# Insulator-metal-superconductor transition in medium-entropy van der Waals compound MEPSe$_3$ (ME=Fe, Mn, Cd, and In) under high pressures

Xu Chen[1†], Junjie Wang[1,2†], Tianping Ying[1*], Dajian Huang[3], Huiyang Gou[3], Qinghua Zhang[1], Yanchun Li[4], Hideo Hosono[5], Jian-gang Guo[1,6*], and Xiaolong Chen[1,6*]

[1] *Beijing National Laboratory for Condensed Matter Physics, Institute of Physics, Chinese Academy of Sciences, Beijing 100190, China;*

[2] *University of Chinese Academy of Sciences, Beijing 100049, China;*

[3] *Center for High Pressure Science and Technology Advanced Research (HPSTAR), Beijing 100094, China;*

[4] *Multidiscipline Research Center, Institute of High Energy Physics, Chinese Academy of Sciences, Beijing 100049, China;*

[5] *Materials Research Center for Element Strategy, Tokyo Institute of Technology, Yokohama 226-8503, Japan;*

[6] *Songshan Lake Materials Laboratory, Dongguan, Guangdong 523808, China*

†These authors contributed equally to this work
*ying@iphy.ac.cn, jgguo@iphy.ac.cn, chenx29@iphy.ac.cn

**Abstract**

MPX$_3$ (M=metals, X=S or Se) represents a large family of van der Waals (vdW) materials featuring with P-P dimers of ~2.3 Å separation. Its electrical transport property and structure can hardly be tuned by the intentional chemical doping and ionic intercalation. Here, we employ an entropy-enhancement strategy to successfully obtain a series of medium-entropy compounds MEPSe$_3$ (ME=Fe, Mn, Cd and In), in which the electrical and magnetic properties changed simultaneously. Lone-pair electrons of P emerge due to the dissociation of the dimers as evidenced by a 35% elongation in the P-P interatomic distance. The band gap widens from 0.1 eV to 0.7 eV by this dissociation. Under external physical pressure up to ~50 GPa, a giant collapse of up to 15% in the *c*-axis happens, which is in contrast to the in-plane shrinkage of their counterparts Fe/MnPSe$_3$. It leads to the recombination of P$^{3-}$ with lone pair electrons into a P-P dimer and the smallest bulk modulus of 28 GPa in MPX$_3$. The MEPSe$_3$ transits from a spin-glass insulator to metal, and to superconductor, which is rarely observed in the MPX$_3$. Our findings highlight the P-P dimer as an indicator to probe diverse electronic structure and the effectiveness of entropy-enhancement in materials science.

**medium-entropy compound, electronic structure, P-P dimer, insulator-metal-superconductor transition, materials science**

# 1 Introduction

Atomic dimers are pairs of atoms connected covalent or metallic bonds, which can adopt or release electrons, acting as a charge reservoir. The elongation or shortening of bond length in a dimer often leads to change in crystal structure accompanying different electron concentrations and magnetism in compounds [1,2]. Typical dimers include Te-Te, O-O, and As-As, as well as metallic bonds Mn-Mn and Ir-Ir. It is reported that the charge redistribution in the interlayer Ir-Ir dimer of $IrTe_2$ is controlled by the interlayer Te-Te dimer, which is believed to be a critical factor for determining charge density wave and superconductivity (SC) [3]. Meanwhile, in Pt- and Pd-doped $IrTe_2$, SC is induced as a result of decoupling the interlayer Te-Te dimers [4,5]. In an O-O dimer, the $\pi^*$ orbital with unpaired electrons endows its magnetic ordering, and diverse $p$-electron magnetic properties are observed in $Rb_4O_6$, and $\alpha$-$BaNaO_4$ due to the differently charged $O_2^-$ or $O_2^{1.5-}$ dimers [6,7,8,9]. In $LaFe_2As_2$, a collapsed-to-uncollapsed phase transition caused by breaking the interlayer As-As dimer is intimate to the emergence of SC [10].

$MPX_3$ (M=divalent metal; X=S, Se) represents a large family of magnetic vdW compounds, which exhibit diverse properties in fundamental and application researches [11,12,13,14,15]. The magnetic $M^{2+}$ ions are honeycombed configuration, thus the isolated monolayer $MPX_3$ is dubbed 'magnetic graphene' with large magnetic anisotropic energy. [16,17,18,19,20]. Apart from transition metals, M ions can be alkaline earth metal ($Ae^{II}$) in $MgPSe_3$, a mixture of +1 ($A^I$) and +3 cations ($RE^{III}$) in $Ag^I In^{III} P_2Se_6$ and partial-occupied $M^{III}$ cation in $(Ga^{III})_{2/3}PSe_3$ [21,22,23,24]. The sum of the chemical valence of M should be +2 so as to keep charge neutrality. SC has long been anticipated in the $MPX_3$ system, but so far, it has only been observed in $FePSe_3$ under external pressure (8 GPa) with a transition temperature ($T_c$) of 4 K [11]. It is proposed that the switch of high-spin to low-spin in Fe $3d^6$ electron induces magnetic-nonmagnetic transition, which is crucial for realizing SC. This explanation is supported by the absence of SC in $MnPSe_3$ with odd $d$-electron number ($3d^5$) under pressure [25]. However, the recent neutron diffraction study on the pressurized $FePSe_3$ reveals that the local magnetic moment survives with ~5 $\mu_B$/Fe in the whole pressure range [26]. The picture of spin crossover may not directly explain the SC in $FePSe_3$.

Despite the variety of M ions, the P-P distance always lies around 2.3 Å, suggesting a robust nature of P-P dimer in $MPX_3$. Field-gating, chemical doping, and molecular intercalation all failed to change the semiconducting or insulating behavior in $MPX_3$ and the reason is not clear so far. Here, we use an entropy strategy to alloy multiple elements in $MEPSe_3$ (ME=Fe, Mn, Cd, and In), which introduces extra electrons leading to breaking the P-P dimer. This is a first report of the control of P-P distance in the $MPX_3$ system with a 35 % elongation from 2.3 to 3.1 Å. Under external pressure, the lone pairs generated from the breaking P-P bonds recombine each other with into P-P dimer, accompanied by an abrupt collapse of the $c$-axis, rather than the $ab$-plane as reported in $FePX_3$ and $MnPX_3$. Accordingly, the $MEPSe_3$ undergoes an insulator-metal-superconductor transition.

## 2 Methods

**2.1 Synthesis.** Single-crystalline medium-entropy MEPSe$_3$ were grown by using the chemical vapor transport (CVT) method. A quartz tube were kept at high temperature of 900 K and low temperature of 650 K. After holding the temperatures for 1 week, the single crystals were obtained after quenching the quartz tube in ice water. I$_2$ was used as the transport agent.

**2.2 Characterization.** Powder X-ray diffraction (PXRD) patterns of samples were collected at room temperature using a Panalytical diffractometer with Cu K$\alpha$ ($\lambda$=1.5408 Å) radiation. The composition of the sample was determined by energy dispersive spectroscopy (EDS). High-resolution transmission electron microscope (HRTEM) and EDS mapping were taken using an ARM-200F (JEOL, Tokyo, Japan) scanning transmission electron microscope (STEM) operated at 200 kV with a CEOS Cs corrector (CEOS GmbH, Heidelberg, Germany) to cope with the probe-forming objective spherical aberration. X-ray photoemission spectrum (XPS) was taken at room temperature. Electrical resistance ($\rho$) and DC/AC magnetic susceptibility ($\chi$) were measured through the physical property measurement system (PPMS, SVS, Quantum Design).

The high-pressure transport property of MEPSe$_3$ was measured in a physical property measurement system (PPMS, Quantum Design) by using a diamond anvil cell (DAC) with a facet diameter of 300 $\mu$m. Four-wire method in a Be-Cu cell was used for the resistance ($R$) measurements. Cubic boron nitride (cBN) powders were employed as the pressure transmitting medium and the insulating material. The pressure was calibrated using ruby fluorescence at room temperature. Synchrotron XRD patterns were collected at the Beamline 4W2 at Beijing Synchrotron Radiation Facility ($\lambda$=0.6199 Å). The beam size were 15 $\mu$m and 5 $\mu$m at the bending and undulator beam-lines, respectively.

## 3 Results and discussion

The methods of sample synthesis and characterizations were put in the supporting infomation. We tried to alloy metals like Fe, Mn, Cr, Co, Zn, Cd, and In at the Fe site of FePSe$_3$, and found only the combination of Fe, Mn, Cd, and In produce single-phase samples. The PXRD patterns and EDS mappings of FePSe$_3$, Fe$_{0.8}$Mn$_{0.1}$Cd$_{0.05}$In$_{0.05}$PSe$_3$ (Fe$_{0.8}$), Fe$_{0.7}$(MnCd)$_{0.1}$In$_{0.1}$PSe$_3$ (Fe$_{0.7}$), and (FeMnCd)$_{0.25}$In$_{0.17}$PSe$_3$ (Fe$_{0.25}$) are presented in Figures S1-S3. Because the ionic radius of Mn (0.67 Å), Cd (0.95 Å), and In (0.80 Å) are larger than that of Fe (0.61 Å), the diffraction peaks systemically shift to lower angles as the content of dopants increases, suggesting that these atoms have been successfully incorporated at the Fe site. All the PXRD patterns of MEPSe$_3$ can be indexed into a trigonal cell with space group $R$-3. Rietveld refinements results of PXRD patterns of MEPSe$_3$ are summarized in Table S1. Figures 1a and b illustrate the crystal structure of MEPSe$_3$, where ME is a mixture of Mn, Fe, Cd, and In. The ME atoms are octahedrally coordinated by six Se atoms that are then covalently bonded to P-P dimers, forming an ethane-like (P$_2$Se$_6$)$^{4-}$ cluster [27,28,29]. In the Fe$_{0.25}$ sample, the ME-ME distance is 3.69 Å, longer than the Fe-Fe of 3.44 Å and Mn-Mn of 3.50 Å in FePSe$_3$ and MnPSe$_3$, respectively [11,25]. The HRTEM image of the Fe$_{0.25}$ sample reveals good crystallinity with partial ME vacancies, see Figure 1c. The EDS mapping of the Fe$_{0.25}$ sample is shown in Figure 1d, indicating a homogeneous elemental distribution of the alloyed elements.

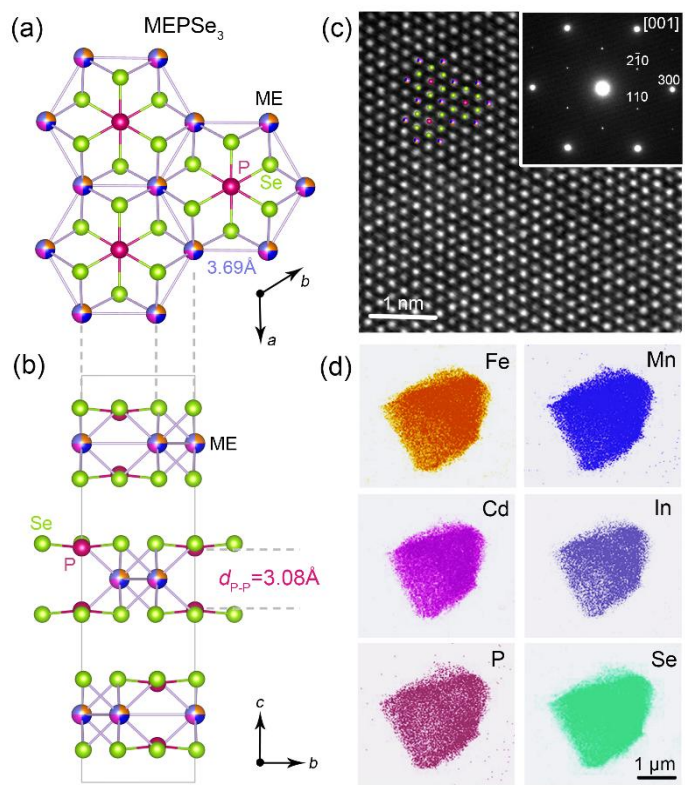

**Figure 1** (Color online) (a, b) Top- and side-view of $(FeMnCd)_{0.25}In_{0.17}PSe_3$ along $c$-axis and $a$-axis. (c) HRTEM image of $(FeMnCd)_{0.25}In_{0.17}PSe_3$ taken along $c$-axis. Inset is indexed SAED image. (d) EDS mapping of Fe (orange), Mn (Blue), Cd (magenta), In (purple), P (brown), and Se (green) of $(FeMnCd)_{0.25}In_{0.17}PSe_3$. The scale bar is 1 μm.

Figure 2 shows the evolution of the structure and physical property of four samples. The lattice constants $a$ and $c$ of four samples are plotted in Figure 2a. By increasing the content of dopants, the $a$-axis increases from 6.26 Å to 6.38 Å, and the $c$ increases from 19.8 Å to 20.0 Å. The expansion of the $c$-axis is mainly caused by the enlargement of P-P distances as shown in Figure 2b. In the $Fe_{0.25}$ sample, the P-P distance is 3.08 Å, which considerably exceeds ~2.3 Å, the typical P-P dimer region. The electrical resistivity versus temperature ($\rho$-$T$ curves) of four samples indicates they are semiconductors as shown in Figure S4. The fitted thermal activated gaps ($E_a$) are plotted in Figure 2b. The $E_a$ increases from 0.1 eV to 0.7 eV. We show the XPS of anion P in Figures 2c. The red-shifts of P's $2p_{5/2}$ and $2p_{3/2}$ peaks indicate that the valences of P qualitatively become more negative. It suggests the formation of partial $P^{3-}$ lone-pair electrons by breaksing of P-P dimers as increasing the dopants.

Temperature-dependent DC magnetic susceptibility ($\chi$-$T$) under zero-field cooling and field cooling modes of $MEPSe_3$ are plotted in Figure 2d and Figure S5. The Neel temperature ($T_N$) of 122 K in $FePSe_3$ is suppressed to a spin-glass-like transition of 15 K in the $Fe_{0.25}$ sample. Fitting the data above $T_N$ by the Curie-Weiss equation yields effective magnetic moments $\mu_{eff}$=5.6 $\mu_B$, 5.5 $\mu_B$, 4.9 $\mu_B$, and 3.7 $\mu_B$ for Fe$_{1.0}$, Fe$_{0.8}$, Fe$_{0.7}$, and Fe$_{0.25}$ samples, respectively. As averaged to magnetic ions (Fe and Mn), the effective magnetic moment is ~5.5 $\mu_B$. With the increase of dopants, the fitted Weiss temperatures ($T_\theta$) change from positive to negative, indicating a transition of local spin interaction from ferromagnetic to antiferromagnetic. We measured the temperature-dependent AC magnetic susceptibility from 10 to 30 K under different frequencies. The peak of $T_r$~15 K shifts to higher temperature with increasing frequency $f$, and the well-fitting results confirm a spin-glass behavior, see Figure S6 [30].

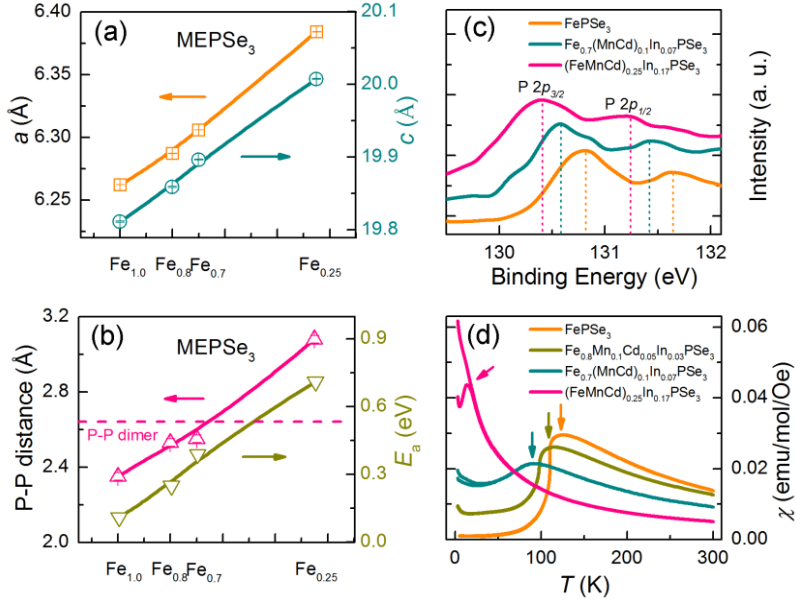

**Figure 2** (Color online) (a) Lattice constants $a$ and $c$ of FePSe$_3$ and three MEPSe$_3$ samples. (b) P-P distance and band gap ($E_a$) estimated from temperature dependence of conductivity in four samples. (c) P 2$p$ XPS of three samples. (d) DC Susceptibility ($\chi$) of four samples as a function of temperature from 2-300 K.

We selected two typical samples, Fe$_{0.8}$ and Fe$_{0.25}$, to study the pressure-induced evolution structure and electrical transport properties. The collected PXRD patterns under pressures are shown in Figures S7 and S8. Figure 3a and 3b present pressure-dependent (113) and (300) peak shifts of Fe$_{0.8}$ and Fe$_{0.25}$. It can be found that both angles monotonously shift to higher values as increasing pressure, indicating shrinkage in the unit cell. At critical pressures of 12 GPa (Fe$_{0.8}$) and 28 GPa (Fe$_{0.25}$), they undergo a kind of quasi-isostructural phase transition, accompanying with an abrupt decrement in lattice constants. To analyze the anisotropic compressibility of FePX$_3$, we refined the PXRD patterns of MEPSe$_3$ using a pseudo-trigonal space group $R$-3 [11,25]. The normalized lattice constants $a$ and $c$ against pressure are plotted in Figures 3c and d. For the Fe$_{0.8}$, the total shrinkage ratios of $a$- and $c$-axis, i.e. $\Delta l=[l_0-l_{(p)}/l_0]*100\%$, are 5.5% and 28.0%, respectively. As crossing the phase transition region, the collapse ratio of $a$- and $c$-axis are 1.3% and 7.7%, respectively. In the Fe$_{0.25}$, the total shrinkage ratios $a$- and $c$-axis are 13.0% and 36.0%, and the collapse ratios of $a$- and $c$-axis are 2.8% and 15.0%, respectively. As previously reported the collapse ratio of $\Delta a$ is ~4%, and the $\Delta c$ is zero in FePSe$_3$. The shrinkage is mostly restricted within *ab*-plane [11]. In our medium-entropy compound MEPSe$_3$, the P-P distance is beyond the covalent bond length. Since the P-P bond is easier compressive along the *c*-axis, the primary effect under compression changes from in-plane to out-of-plane. Notewhrthy is that the giant shrinkage of 15% can lead to the recombination of P-P dimer.

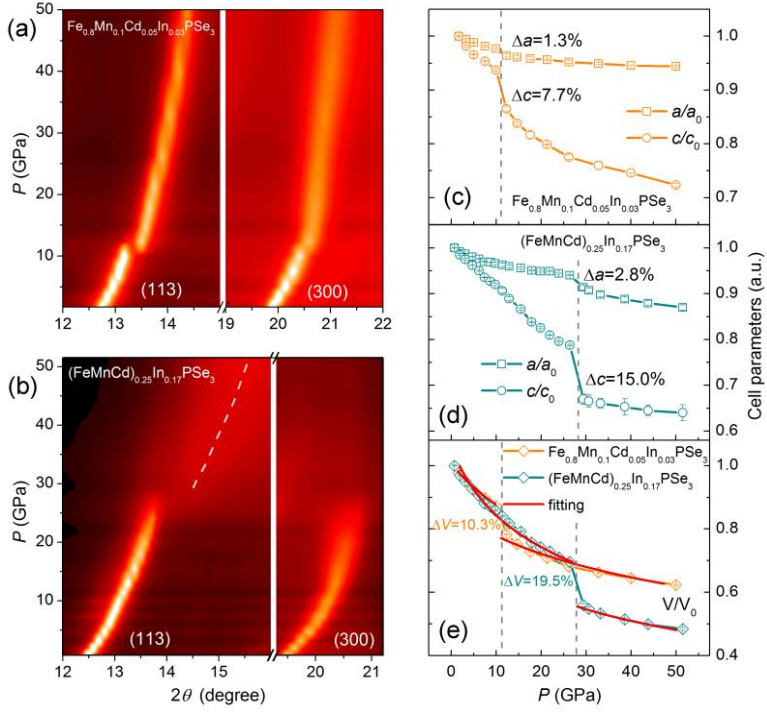

**Figure 3** (Color online) (a, b) Color contour of the (113) and (300) diffraction peaks for Fe$_{0.8}$Mn$_{0.1}$Cd$_{0.05}$In$_{0.05}$PSe$_3$ (Fe$_{0.8}$) and (FeMnCd)$_{0.25}$In$_{0.17}$PSe$_3$ (Fe$_{0.25}$) under external physical pressure. (c-e) Pressure-dependent lattice parameters and volume of unit cell for both samples.

Figure 3e displays the plot of unit-cell volume (*V*) of Fe$_{0.8}$ and Fe$_{0.25}$ samples under pressure. The shrinkage ratios of *V* from ambient pressure to 50 GPa are 38% and 52%, and the collapse ratios at the phase transition pressure are 10.3% and 19.5%, respectively. We fitted the pressure-dependent *V* of both samples by using the third-order Birch-Murnaghan equation of state: [31,32,33]

$$P = \frac{3}{2}B\left[(V_0/V)^{7/3} - (V_0/V)^{5/3}\right]\left\{1 + \frac{3}{4}(B' - 4)\left[(V_0/V)^{2/3} - 1\right]\right\}$$

where *B* is the average isothermal bulk modulus, *B'* the first pressure derivative of *B*, *V* and *V$_0$* the volume under pressure and zero-pressure volume. The fitted $B_{LP}$ ($B_{HP}$) for the low (high)-pressure phase Fe$_{0.8}$ and Fe$_{0.25}$ samples are 55(78) GPa and 27(28) GPa, respectively. The values of Fe$_{0.25}$ is the smallest B ever reported in MPX$_3$ compounds [11]. In general, the *B* of a high-entropy alloy is higher than that of its individual components [34]. Based on our results, the *B* in layered compounds may be mainly determined by the strength and length of chemical bonds. Here, we think that the existence of soft P-P covalent bonds drastically reduces the *B*.

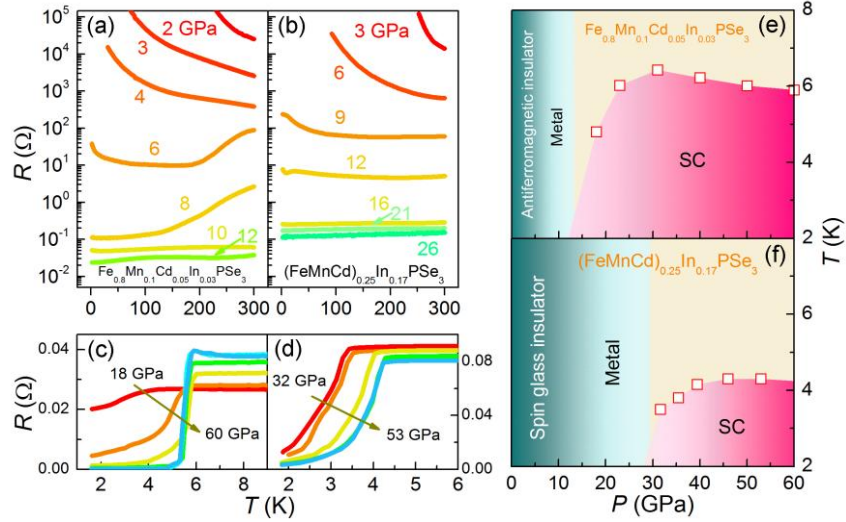

**Figure 4** (Color online) (a, c) Temperature dependence of $R$-$T$ curves of $Fe_{0.8}Mn_{0.1}Cd_{0.05}In_{0.05}PSe_3$ below and above 12 GPa. (b, d) Temperature dependence of $R$-$T$ curves of $(FeMnCd)_{0.25}In_{0.17}PSe_3$ below and above 26 GPa. (e, f) Electronic phase diagram of both samples versus pressure.

In Figures 4a and 4b, we can see that the resistance ($R$) is drastically reduced by an order of $10^6$ in magnitude with pressure. The semiconductor-metal transitions of $Fe_{0.8}$ and $Fe_{0.25}$ occur at 6 and 16 GPa, and the metallic state persists in the pressure range of 6-18 GPa and 16-32 GPa, respectively. The superconducting transitions in $Fe_{0.8}$ and $Fe_{0.25}$ are observed at $T_c$=4 K and 3.2 K, respectively, at pressures above 18 and 32 GPa, see Figures 4c and 4d. These pressures are consistent with the critical pressures of structural collapse shown in Figures 3c and 3d. The maximal $T_c$ of the $Fe_{0.8}$ sample is 6.2 K and then gradually decreases. For the $Fe_{0.25}$ sample, the $T_c$ increases to 4.3 K at 46 GPa and then saturates. The upper critical fields $\mu_0 H_{c2}(0)$ are 1.7 T for the $Fe_{0.8}$ (50 GPa) and 2.9 T for the $Fe_{0.25}$ (53 GPa) according to Ginzburg-Landau fitting, see Figures S9. Their superconducting coherence lengths ($\xi$) are calculated to be 439 Å and 336 Å, respectively. Noted that the SC originating from a spin-glass insulator is a rare case, which deserves detailed investigations on electronic structure.

$FePSe_3$ undergoes an in-plane collapse under external pressure, forming a pseudo-trigonal monoclinic structure with intralayer Fe-Fe bonding [11]. This is reminiscent of the in-plane formation of Ir-Ir dimer in $CdI_2$-type $IrTe_2$, where the charge redistribution between Ir-Ir and Te-Te regulates the CDW and SC [3]. In pyrite-type $IrSe_2$ and $IrTe_2$, on the other hand, the elongated Te-Te/Se-Se distances are responsible for the emergence of SC [35,36]. Similar behavior has been reported in the collapsed to unclasped transition in $LaFe_2As_2$ [10].

The status of the P-P dimers has also been reported in tuning magnetism. In $Sr_{1-x}Ca_xCo_2P_2$, the breaking of the P-P dimer is closely related to a transition from magnetic ($x$=1) to nonmagnetic ($x$=0) [37]. In $SrCo_2(Ge_{1-x}P_x)_2$, a ferromagnetic quantum critical point has been discovered on the verge of Ge(P)-Ge(P) dimer instability [38]. In our work, under high pressure, the collapse of the lone pair of P triggers a transition from $(P_2)^{6-}$ to $(P-P)^{4-}$ and releases a certain amount of electrons. This charge redistribution significantly alters the electronic structure, *i. e.* increasing the carriers concentration and density of states at the Fermi level, in favor of SC [39]. To our knowledge, MEPSe$_3$ is the first case that SC is directly linked to the status of the P-P dimers.

## 4 Conclusion

The P-P dimers are extremely flexible upon electron doping and under high pressure, which can act as a charge reservoir. We adopt a novel strategy by alloying multiple elements in medium-entropy MEPSe$_3$ compounds, in which breaking P-P dimers increases the electrical resistivity and activiated gap. Under external physical pressure, the P-P dimes recombine as evidenced by the dramatic collapse of the *c*-axis, which induces SC at the same time. It provides an effective way to tune crystalloghpic and electronic structure of MPX$_3$. Our findings highlight the importance of synergistic entropy-driven and high-pressure effect in materials science.

*This work is financially supported by the National Key Research and Development Program of China (No. 2018YFE0202600, 2017YFA0304700, and 2021YFA1401800), the National Natural Science Foundation of China (No. 51922105) and Beijing Natural Science Foundation (Grant No. Z200005).*

**Supporting Information**

The supporting information is available online. The supporting materials are published as submitted, without typesetting or editing. The responsibility for scientific accuracy and content remains entirely with the authors